\documentstyle[12pt,epsfig]{ioplppt}
\newcommand {\be} {\begin{equation}} 
\newcommand {\ee} {\end{equation}} 
\newcommand {\Be}{\begin{eqnarray*}}
\newcommand {\Ee} {\end{eqnarray*}}
\newcommand {\bey} {\begin{eqnarray}} 
\newcommand {\eey} {\end{eqnarray}}

\newcommand{\new}{}

\begin{document}
\jl{1}
\title{Mean--field theory of critical coupled map lattices}[Mean--field 
theory of coupled map lattices]
\author{Stefano Lepri\footnote{e--mail: lepri@mpipks-dresden.mpg.de} 
and Wolfram Just\footnote{e--mail:
wolfram@mpipks-dresden.mpg.de}}
\address{Max--Planck Institute for Physics of Complex Systems,
N\"othnitzer Str.~38, D--01187 Dresden, Germany}
\vspace{0.5cm}
\begin{abstract}
We {\new study} the single--site approximation of the Perron--Frobenius 
equation for 
a coupled map lattice exhibiting a phase transition at a critical value $g_c$ 
of the coupling constant. We found that the critical exponents are the same as 
in the usual mean--field theory of equilibrium statistical mechanics. 
Remarkably, 
the value of $g_c$ is within six percent of the one previously obtained by 
numerical simulations with asynchronous updating.
\end{abstract}
\maketitle
\section{Introduction}
Critical phenomena in extended dynamical systems attracted considerable 
interest recently. {\new In particular, coupled map lattices were 
introduced as 
paradigmatic models of nonequilibrium systems undergoing (in the large--size 
limit) a second--order phase transition between two chaotic states 
\cite{Sakaguchi,milhu}. In those models such a transition is associated with 
a breaking of the Ising symmetry, and is thus expected to share the same static 
critical exponents of the Ising model itself. However, numerical simulations 
\cite{chate} show that this is not the case. More precisely, the critical 
exponent of the correlation length for the model of Ref.~\cite{milhu} turned 
out to be significatively different from the Ising value. Moreover, the latter 
are recovered as soon as the updating rule is changed from parallel to 
sequential, thus indicating that some features of the microscopic dynamics may 
be relevant for the critical behaviour. Besides of those facts, it is not 
completely clear why a phase transition behaviour emerges at all and why the 
coupling is effectively ``ferromagnetic''.

For all the above reasons, even the simplest analytical approach, namely the 
mean--field like approximation, is worth to be investigated. Actually, it is 
not clear to which extent it can be applied in nonequilibrium cases and 
which properties it may share with its equilibrium counterpart. Moreover, it 
is still questionable that such an approach is able to reproduce the 
qualitative behaviour of coupled map lattices in general.

We will discuss such problems for the coupled map lattice introduced in Ref. 
\cite{milhu}. Its equations of motion read as
\be
x_{n+1}^{(\nu)} = (1-gd)f(x_n^{(\nu)}) + g \sum_{\mu\in {\cal U}_{\nu}}
f(x_n^{(\mu)}) \quad ,
\label{cml}
\ee
where $x_n^{(\nu)}$ is the field variable at time $n$ and the
index $\nu$ enumerates the sites on a square lattice. Here, ${\cal U}_{\nu}$ 
denotes the set of the $d$ neighbours to which the $\nu$th site is coupled 
and $g$ is the coupling constant.
Rather than considering the original isotropic nearest--neighbour coupling 
on a square lattice we will refer to its modification presented 
in \cite{chate},
which couples four instead of five sites. This choice has the advantage of 
simplifying the algebraic calculations and offers the possibility to compare 
our results with the numerical simulations reported in \cite{chate}. Since we 
employ some mean--field theory, our approach applies regardless of the special 
type of spatial geometry and boundary conditions do not play a 
significant role.} 

The local map, which is defined on the interval $[-1,1]$
\be \label{singmap}
f(x) = \cases{ 
-3x-2  \quad & for $-1\le x\le -1/3$ \cr
3x \quad &for $-1/3 < x < 1/3$ \cr
-3x+2 \quad &for $1/3 \le x\le 1$ 
}
\ee
obeys an Ising like symmetry $f(x)=-f(-x)$. In the supercritical region 
$g>g_c$ the symmetry is spontaneously broken and the coupled map lattice 
exhibits two ordered phases, characterised by a non--vanishing value of 
the ``magnetisation'' $\langle \sum_{\nu} x^{(\nu)}\rangle$. 

\section{Mean--field approach}
Our starting point is the evolution equation for the single--site probability
density $\rho(x)$ (cf.~e.~g.~\cite{ruelle}), 
{\new as usually obtained by projection of the Perron--Frobenius
equation for the full probability density. The single--site density}
is expressed in terms of the four--sites joint probability density 
$\rho^{(4)}$ as
\be
\rho_{n+1}(x) = \int dy_0 dy_1 dy_2 dy_3 
\delta\left[ x-T(y_0,y_1,y_2,y_3)\right]
\rho_n^{(4)}(y_0,y_1,y_2,y_3) 
\label{fp} \quad .
\ee
Here we have introduced the abbreviation
\be 
T(y_0,y_1,y_2,y_3) = (1-3g)f(y_0) + g\left[f(y_1)+f(y_2)+f(y_3)
\right] \quad .
\ee
The function $T$ is invariant under all permutations of the variables 
$y_1,y_2,y_3$. The mean--field Perron--Frobenius equation \cite{kaneko} 
is as usual obtained  by neglecting multiple correlations, namely by 
letting
\be
\rho_n^{(4)}(y_0,y_1,y_2,y_3) \;\rightarrow\; \rho_n(y_0)
\rho_n(y_1)\rho_n(y_2)\rho_n(y_3) \quad .
\label{fac}
\ee
Then eq.(\ref{fp}) becomes a nonlinear integral equation
\be
\fl
\rho_{n+1}(x) =\int dy_0 dy_1 dy_2 dy_3 
\delta\left[ x-T(y_0,y_1,y_2,y_3)\right]
\rho_n(y_0)\rho_n(y_1)\rho_n(y_2)\rho_n(y_3) 
\label{mffp} \quad .
\ee
It should be stressed that the approximation (\ref{fac}) leads to different 
results if one chooses different coordinate systems in the full phase space 
of the map lattice. In fact, neglecting correlations has a different meaning
in different coordinate systems. Hence, similar to statistical mechanics, 
it makes no sense to speak about {\it the} mean--field approximation. The 
formulation chosen here seems to be quite appropriate for the analytical 
calculation.

Since we are interested in stationary properties we look for the 
fixed point solution $\rho_n=\rho_*$. In the paramagnetic region, 
i.~e.~for sufficiently small coupling, the solution is symmetric
$\rho_*(x)= \rho_*(-x)$ and does not give rise 
to a finite magnetisation, 
namely  $\int dx \, x \rho_*(x) =0$. At the critical coupling $g_c$ such 
solution will get unstable in favour of a non--symmetric one.
In order to tackle this problem we first solve for the symmetric stationary 
solution employing a Fourier series expansion
\be
\rho_*(x) = \frac{1}{\sqrt{2}} \sum_{k=-\infty}^{+\infty}
c_{k}\, \exp(i \pi k x)
\quad .
\label{fourier}
\ee
Here, because of normalisation $c_{0}=1/\sqrt{2}$, and all the 
coefficients $c_{k}= c_{-k}$ are real because of symmetry. 
With the abbreviation
\bey
F_{k k'}(g) &=& \int \exp[i \pi (k' x - k g f(x))] \, dx \nonumber \\
&=& 4 \cos\left(2 \pi k'/3\right) \frac{\sin\left[\pi(k'+3 k g)/3\right]}
{\pi(k'+3 k g)} + 2 \frac{\sin\left[\pi(k'-3 k g)/3\right]}
{\pi(k'-3 k g)}
\label{mat}
\eey
the fixed point equation reads
\be
c_{k}= \frac{1}{2^{5/2}}
\left( \sum_{k'} F_{kk'}(1-3g) c_{k'} \right)
\left( \sum_{k''} F_{kk''}(g) c_{k''} \right)^3
\label{fix}
\quad .
\ee
Although such an equation can be numerically solved by iterative methods, 
it is convenient to look for an approximate analytical solution. 
As $F_{k k'}+F_{k -k'}=0$ holds whenever the index $k'$ is not an
integer multiple of three, the right hand side of equation (\ref{fix})
contains only Fourier coefficients of the form $c_{k'=3 l}$ for
symmetric densities.  
If we truncate the expansion (\ref{fourier}) at the fifth mode, then the right 
hand side of eq.(\ref{fix}) only contains the coefficient $c_{3}$.
The latter is self consistently determined by
\bey
\sqrt{2}c_{3} &=& \frac{1}{4}
\left(F_{3 0}(1-3 g)+[F_{33}(1-3g)+F_{3 -3}(1-3g)]
\sqrt{2} c_{3}\right) \nonumber\\
 & & \left(F_{3 0}(g)+[F_{33}(g)+F_{3 -3}(g)]
\sqrt{2} c_{3}\right)^3 
\quad . \label{selfcon}
\eey
The polynomial (\ref{selfcon}) admits two real solutions in the whole range 
$0<g<1/3$, but only one of them is smaller than $1/2$. The remaining Fourier 
coefficients up to order five are now obtained if we plug in the expansion 
(\ref{fourier}) with the solution of eq.(\ref{selfcon}) into the right hand 
side of eq.(\ref{fix}). The analytical expression thus obtained coincides up 
to five digits with the full numerical solution of eq.(\ref{fix}) if
many modes are taken into account. 

We are now going to evaluate explicitly the critical point. It 
corresponds to the value of coupling where the symmetric solution loses 
its stability. Considering therefore small deviations from it 
$\rho_n= \rho_*+\delta \rho_n$, and expanding eq.(\ref{mffp}) 
we obtain
\be \label{dr}
\delta\rho_{n+1}(x) = ({\cal L} \delta \rho_n)(x) +
({\cal C} [\delta \rho_n,\delta \rho_n])(x) +
({\cal D} [\delta \rho_n,\delta \rho_n,\delta \rho_n])(x) +
\cdots \quad .
\ee
The deviations obey the constraint $\int dx\, \delta \rho_n =0$ because of the
normalisation condition. The stability properties are determined by the 
eigenvalue problem for the linear operator
\be
({\cal L} \delta \rho)(x)\; = \; \int dy \, \Lambda(x,y) \delta \rho(y)
\quad .
\label{op}
\ee
Its kernel reads
\bey
\Lambda(x,y)&=&\int \, dz_1 dz_2 dz_3 \, \left(
\delta\left[x-T(y,z_1,z_2,z_3)\right] 
+3\delta\left[x-T(z_1,y,z_2,z_3)\right] \right) \nonumber \\
& &
\rho_*(z_1)\rho_*(z_2)\rho_*(z_3)
\label{ker}
\quad .
\eey
Under quite mild conditions on $\rho_*$ (e.~g.~continuity is 
sufficient) the kernel is continuous. Hence the operator (\ref{op})
is compact and its spectrum consists of isolated eigenvalues which 
accumulate at most at zero \cite{funct}. In addition, the bifurcation 
behaviour of the symmetric density is essentially identical to bifurcations 
in low--dimensional dynamical systems. In particular, the instability is 
typically caused by a single eigenvalue $\lambda$ crossing the unit circle 
in the complex plane.

\begin{figure}
\begin{center}
\mbox{\epsfig{figure=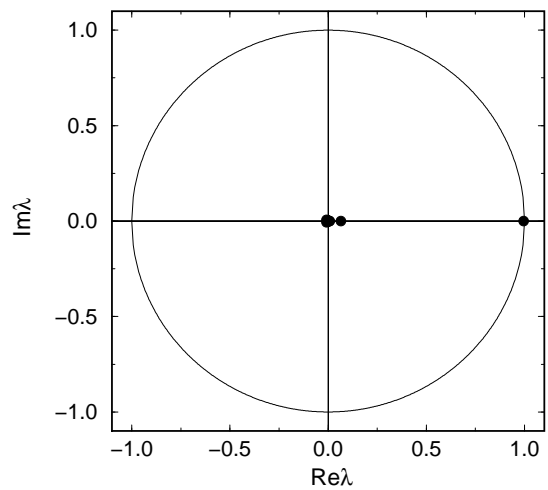,width=9cm}}
\end{center}
\caption[ ]{Spectrum of the operator (\ref{op}) at the critical
coupling $g_c$. \label{fig1}}
\end{figure}

\begin{figure}
\begin{center}
\mbox{\epsfig{figure=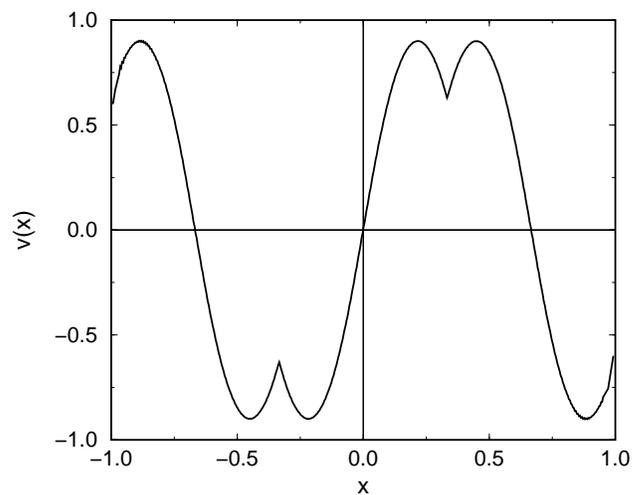,width=9cm}}
\end{center}
\caption[ ]{The critical eigenfunction $v(x)$ corresponding to
$\lambda=1$. \label{fig2}}
\end{figure}

The numerical solution of the eigenvalue problem is accomplished by 
representing $\Lambda$ on a truncated Fourier basis, and diagonalising
the resulting finite--dimensional matrix. The spectrum thus obtained at the 
critical coupling is displayed in figure \ref{fig1}. An isolated eigenvalue 
attains the unit circle along the real axis and the corresponding critical 
eigenfunction $v(x)$ is odd with respect to the space inversion $v(x)=-v(-x)$ 
(cf.~figure \ref{fig2}) 
\footnote{Since the full equation (\ref{mffp}) is homogeneous of order four
the linear operator (\ref{op}) admits a Goldstone--like mode with eigenvalue 
$\lambda=4$ and eigenfunction $\rho_*(x)$.}.
Notice that, as the kernel (\ref{ker}) is not symmetric and the corresponding 
linear operator (\ref{op}) is in general not selfadjoint, the critical 
left--eigenfunction $w(x)$ may differ from $v(x)$. 
From the numerical diagonalisation of the matrix the critical coupling is 
very accurately determined to be $g_c=0.1496\ldots$, which remarkably 
deviates only 6\% from the value $0.1584\ldots$ of the critical point obtained 
in numerical simulations of the full map lattice with asynchronous updating 
\cite{chate}. This coincidence is consistent with the idea that the latter 
dynamic rule induces less spatial correlations than the synchronous one.

\section{Bifurcation analysis}

{\new The bifurcation scenario described at the end of the preceding 
section is of the pitchfork type. Hence the growth of the critical mode 
beyond the instability is expected to scale as $(g-g_c)^{1/2}$. We stress that 
such a critical behaviour follows solely from the compactness of the linear 
operator. Therefore, we expect that every mean--field approximation of the
Perron--Frobenius equation shares the same critical exponents. 

To support
such general remarks we are now going to perform
a complete normal form reduction} of
the full mean--field equation. For that purpose one needs the quadratic 
and cubic contributions {\new of eq.(\ref{dr})} which are given by
\bey
({\cal C}[\varphi,\psi])(x) &=& \int dy_1 dy_2 
\Gamma(x,y_1,y_2)
\varphi(y_1) \psi(y_2) \label{cterm}\\
({\cal D}[\varphi,\psi,\omega])(x) &=&
\int dy_1 dy_2 dy_3 \Delta(x,y_1,y_2,y_3) \varphi(y_1) \psi(y_2) \omega(y_3)
\label{dterm}
\quad .
\eey
The corresponding kernels read for example 
\bey
\fl
\Gamma(x,y_1,y_2) &=& \int \, dz_1 dz_2 \left[ 3 \delta(x-T(y_1,y_2,z_1,z_2)) +
3 \delta(x-T(z_1,y_1,y_2,z_2))\right]\nonumber \\ 
\fl & & \rho_*(z_1)  \rho_*(z_2) \\
\fl
\Delta(x,y_1,y_2,y_3) &=& \int \, dz \left[ \delta(x-T(z,y_1,y_2,y_3))+
3 \delta(x-T(y_1,z,y_2,y_3))\right] \rho_*(z) \quad .
\eey
The centre manifold tangential to $v(x)$ is expanded as
\be
\delta \rho_n (x) = \alpha_n v(x) + \alpha_n^2 r(x) + \alpha_n^3  
s(x) +\cdots \quad ,
\label{cm}
\ee
where the scalar $\alpha_n$ denotes the coordinate on the one dimensional
manifold. To remove the ambiguity on the transversal vectors $r(x)$ and 
$s(x)$, we chose $\langle w| r\rangle=0$, $\langle w | s \rangle=0$ with 
respect to
the canonical bilinear form $ \langle \psi | \varphi \rangle = \int dx 
\, \psi(x) \varphi(x)$.  The time evolution on the centre manifold obeys
\be \label{cmev}
\alpha_{n+1}= A\alpha_n + B \alpha_n^2 + C \alpha_n^3+ \cdots \quad .
\ee
If we now plug in eqs.(\ref{cm}) and (\ref{cmev}) into eq.(\ref{dr}) and
compare the different powers in $\alpha_n$, then the unknown expansion
coefficients in eq.(\ref{cmev}) will be fixed. To the first order we just
obtain the eigenvalue equation
\be
({\cal L} v)(x)  = A v(x)
\ee
so that $A=\lambda$. To the second order we obtain a linear 
inhomogeneous equation determining the transversal vector $r$
\be \label{sec}
A^2 r(x) - ({\cal L} r)(x)  = ({\cal C}[v,v])(x) 
-B v(x) \quad .
\ee
Since the operator on the left hand side becomes singular at the
critical point $g=g_c$, but the solution has to stay regular, the right hand
side must obey a Fredholm condition at $g=g_c$. In particular,
the right hand side has to be orthogonal to $w(x)$. Moreover, as the
first term on the right hand side drops from this condition by symmetry, 
we are left with $B=0$. Finally, at the third order we obtain
\be \label{thi}
\fl
A^3 s(x) - ({\cal L} s)(x) =
({\cal C}[v,r])(x) +({\cal C}[r,v])(x)
+ ({\cal D} [v,v,v])(x) - C v(x) \quad .
\ee
The Fredholm condition ensuring a regular solution at criticality
determines the cubic coefficient as
\be \label{koe}
C\langle w | v \rangle =
\langle w | {\cal C}[v,r] +{\cal C}[r,v] 
+ {\cal D} [v,v, v] \rangle \quad .
\ee
Eq.(\ref{koe}) is easily evaluated using eqs.(\ref{cterm}), (\ref{dterm}),
and (\ref{sec}). If the representation in terms of Fourier modes is
employed, then all matrix elements can be expressed in terms of
(\ref{mat}). Of course the modulus of $C$ has no special meaning, since it
depends on the normalisation of $v(x)$. We just chose $\int dx \, v^2(x)=1$ 
and obtain $C=-8.889\times 10^{-3}$. Hence, the pitchfork bifurcation is 
supercritical as already mentioned above. 
{\new The stationary distribution in the supercritical region is thus 
readily evaluated from the normal form (\ref{cmev}) as
$\rho_*(x) + \sqrt{(\lambda-1)/(-C)} \, v(x)$, where 
$\rho_*$ is the
(symmetric) stationary density at the critical point $g_c$.
Accordingly, the critical mode and the magnetisation grow like 
$(\lambda-1)^{1/2}\simeq (g-g_c)^{1/2}$ beyond criticality.
In addition, it is important to notice that the system is close to a 
super--subcritical transition. In fact,} if one considers the definition 
(\ref{koe}) for arbitrary coupling $g$, 
then a change in the sign of $C$ occurs slightly below the
transition point $g_c$ at $g=0.1477\ldots$. For that reason, 
the correct scaling behaviour occurs only in a narrow region beyond
$g_c$, and the evaluation of the critical exponents from a numerical 
solution of eq.(\ref{mffp}) is almost impossible to perform. 
The last point again emphasises the importance of our analytical 
results.
 
{\new Sofar we have dealt with the critical behaviour of the order 
parameter only. Let us discuss now the counterpart of the static 
susceptibility in equilibrium systems. In analogy with the latter, we 
may tentatively define it as the derivative of the order parameter with 
respect to a suitable ``symmetry breaking field''. This amounts to introduce 
some external parameter $h$ spoiling the symmetry of the 
original single site map. Although not unique, a natural choice is for example 
to replace $f$ in eq.(\ref{cml}) with
\be \label{assym}
f_h (x) = \cases{ 
-3x-2  \quad & for $-1\le x\le -1/3$ \cr
3x \quad &for $-1/3 < x < 1/3$ \cr
-3(1+h) x+2-h \quad &for $1/3 \le x\le 1$ } \quad.
\ee
Here the transition rate from $[0,1]$ to $[-1,0]$ is lowered upon
increasing the parameter $h$, thus mimicking the effect of a
static magnetic field in the Ising system. The mean--field susceptibility 
can then be defined as
\be \label{sus}
\chi\; := \; \left . \frac{\partial \langle x \rangle}{\partial h}
\right|_{h=0} 
\;=\; \left . \frac{\partial}{\partial h} 
\int dx \,x \,\rho_*(x)  \right|_{h=0} \quad , 
\ee
where 
$\rho_*(x)$ denotes now the fixed point solution of eq.(\ref{mffp}) with the
single site map (\ref{assym}).
If we suppose that such a solution depends on $h$ in a differentiable way, 
we can take the formal derivative of eq.(\ref{mffp}) obtaining
\bey \label{difffp}
\fl 
(1- {\cal L})\left.\frac{\partial \rho_*}{\partial h}\right|_{h=0}(x) 
&=&
-\int dy_0 dy_1 dy_2 dy_3 \delta'[x-T_{h=0}(y_0,y_1,y_2,y_3)] \nonumber \\
\fl & &
\left.\frac{\partial T_h(y_0,y_1,y_2,y_3)}{\partial h}\right|_{h=0}
\rho_*(y_0) \rho_*(y_1) \rho_*(y_2) \rho_*(y_3) \quad,
\eey
where the definition of (\ref{op}) has been also taken into account. In order 
to solve this equation for $\partial \rho_*/\partial h|_{h=0}$, we 
expand both the latter quantity and the the right hand side of 
eq.(\ref{difffp})
in terms of eigenmodes of ${\cal L}$. As only one eigenvalue crosses
the unit circle and the right hand side of eq.(\ref{difffp}) remains 
bounded under quite mild conditions on $\rho_*$, all the coefficients 
of such expansion remain bounded in the vicinity of the critical point, 
except for the one of the critical mode $v(x)$. This coefficient
develops a singularity of the form $|\lambda -1|^{-1}$,
which carries over to $\partial \rho_*/\partial h|_{h=0}$, provided 
that the expansion in terms of eigenmodes converges absolutely.

The same result can of course be derived along the lines of bifurcation 
theory. 
Since the instability without the symmetry breaking field is governed by the 
pitchfork normal form and only a single eigenvalue becomes critical, one expects 
from the very beginning that the symmetry breaking unfolds the normal form to 
the cusp case (cf.~\cite{guho}). That indeed occurs if one follows the normal 
form reduction presented above. Altogether, we can conclude that the static 
susceptibility diverges as $|g-g_c|^{-1}$, in accordance with the simple 
mean--field theory for equilibrium systems.}

\section{Conclusions}

We have shown that a mean--field approach for the Miller--Huse
model reproduces the critical behaviour known from simple mean--field 
approximations in equilibrium statistical mechanics. The value of the 
critical exponents basically originates from the compactness property
of the linearised operator. Since such a property holds quite generally, 
our result is supposed to apply for a large class of coupled map lattices and 
almost all kinds of mean--field approximations. Corrections to mean--field 
scaling can only come from a continuous spectrum of the full Perron--Frobenius 
operator, similar to findings in equilibrium statistical mechanics, e.~g.~for 
the two dimensional Ising model. The coincidence with equilibrium mean--field 
theories is far from being obvious. For instance, globally coupled maps behave 
quite different compared to the mean--field approximation of the type 
employed here (cf.~\cite{me} and references therein). That observation is 
in striking contrast to equilibrium statistical mechanics, where mean--field 
approaches and long--range coupled models are often equivalent.

{\new

As mentioned in the beginning, the critical behaviour of the full
map lattice depends on the updating rule. In particular, the sole occurrence
of equilibrium Ising exponents for asynchronous updating was
attributed to a kind of destruction of coherence by the updating rule.
Of course, the plain mean--field approach cannot explain such differences 
in critical exponents at all. Nevertheless, the mentioned interpretation
is fully consistent with the fact that the mean--field approach (which 
neglects all correlations) yields a good estimate of $g_c$ for the model 
with asynchronous updating. 

At equilibrium, the static susceptibility can be expressed in terms 
of the spatial correlation function by virtue of the properties of the 
canonical distribution. This is of course no longer true for out of 
equilibrium 
systems as the coupled map lattices. Hence, it would be tempting to check 
whether even in this case the critical behaviour of the susceptibility 
coincides 
with that of spatial correlations. Such a coincidence, which of course 
requires 
quite accurate numerics, would indicate a relation between response and 
correlations in nonequilibrium systems too.
}

\end{document}